\documentclass{sf2a-conf2011}
\usepackage{graphicx}
\usepackage{hyperref}
\usepackage[]{natbib}  
\usepackage[cyr]{aeguill}
\usepackage{epstopdf}

\def\BibTeX{{\rm B\kern-.05em{\sc i\kern-.025em b}\kern-.08em
    T\kern-.1667em\lower.7ex\hbox{E}\kern-.125emX}}
\bibpunct{(}{)}{;}{a}{}{,}  %%%%%%%%%%%%%  A&A bibliography style
\begin{document}

\TitreGlobal{SF2A 2011}

%%-----------------------------------------------------------------
%%      the top matter
%%

\title{Seismic analysis of two solar-type stars observed by {\it Kepler}}

\runningtitle{ Seismic analysis of two solar-type stars}

\author{S. Mathur}\address{High Altitude Observatory, NCAR, P.O. Box 3000, Boulder, CO 80307, USA}

%% Keep this line, even if the page will be settled afterwards.
\setcounter{page}{237}

%% To make the final index, repeat the authors here, in the format : Surname, Initial(s) 
\index{Mathur, S.}
\index{Campante, T.~L.}
\index{Handberg, R.}

%%-----------------------------------------------------------------

\maketitle

%%-----------------------------------------------------------------
%%        The abstract
%% 
%%  Warning!  within the abstract:
%%  - do not use macros. 
%%  - do not use commands like: \cite, \citet, \citep ... etc.

\begin{abstract}
After more than one year of operation, the Kepler photometer has already provided exquisite data of solar-type stars. During the survey phase, 42 stars have been continuously  observed. It appeared that five stars show evidence of oscillations, even though they are rather faint (magnitudes from 10.5 to 12).
We will show the results of the seismic analysis of the light curves of two of these stars, which have been observed during more than 8 months. This analysis led to the determination of the acoustic-mode global parameters (mean large separation, mean small separation...), lists of frequencies built by comparing the results of several teams, some parameters of the modes, and the rotation period of the stellar surface.
\end{abstract}

%% Insert the keywords (to appear in the ADS indexing)
%% Keywords must be separated by a comma

%%-----------------------------------------------------------------

\section{Introduction}
%%---------------------
Interferometry, spectroscopy, and 
spectro-polarimetry are the classical methods to characterize stars, 
giving us information about mass and radius, the stellar atmosphere, 
rotation and magnetic activity. Over the past decade, {\it 
asteroseismology} has emerged as a powerful new tool allowing us to 
directly probe the stellar interior.

A star is a resonant cavity where two types of waves can propagate. 
Acoustic (p) modes are excited in the convective zone and sustained by 
pressure. These modes are equally spaced in frequency and we define the mean large frequency separation, $\langle \Delta \nu \rangle$, that is the mean root square of the density. The mean small frequency separation, $\langle \delta_{02} \rangle$, is very sensitive to the structure of the star, and thus to its age. Gravity (g) modes propagate in the radiative zone and are 
sustained by buoyancy, rendering them evanescent in the convective zone. 
As such, g modes in solar-type stars are very difficult to detect, since 
they reach the stellar surface with very low amplitudes \citep{2007Sci...316.1591G,2010A&ARv..18..197A}. Finally there are 
mixed modes, which are sustained by both pressure and buoyancy, making 
them sensitive to both the core and the envelope of a star. These modes 
are very precious, as they carry information on the stellar core and reach 
the stellar surface with larger amplitudes than pure g modes \citep[e.g.][]{2011Sci...332..205B,2011Natur.471..608B,2011A&A...532A..86M}.
  
The ground-based observing campaigns \citep[e.g.][]{2008ApJ...687.1180A} and space-based photometric observations with instruments such as WIRE \citep[Wide-Field Infrared Explorer][]{2007A&A...461..619B} and MOST \citep[Microvariability and Oscillations of STars][]{2003PASP..115.1023W} allowed us to study these solar-like oscillations in a few tens of stars. The CoRoT mission \citep[Convection, Rotation, and Transits][]{2006cosp...36.3749B} increased the number of main-sequence solar-like pulsators known thanks to longer continuous observations. Depending on the signal-to-noise ratio, it was possible to estimate the global parameters of the p modes \citep{2009A&A...506...41G,2009A&A...506...33M,2010A&A...518A..53M} and even resolve the individual modes \citep[e.g.][]{2011A&A...530A..97B}.
  
 The {\it Kepler} mission \citep{2010Sci...327..977B}, launched in March 2009 into an Earth-trailing orbit, was designed to search for Earth-like exoplanets. It is a photometer composed of 42 CCDs (now 38) that will monitor the brightness of more than 150,000 stars in the direction of Cygnus and Lyra for at least 3.5 years. An additional program was introduced to apply asteroseismic methods to the target stars. Around 2000 solar-type stars have been observed for one month during the first year of survey and solar-like oscillations were detected in around 600 stars. With this large number of stars, we can start to do ``ensemble'' asteroseismology leading to very interesting results \citep{2011Sci...332..213C}, which are important for our understanding of stellar evolution. Among the stars continuously observed during the survey phase, five present solar-like oscillations: two F-type stars (KIC~11234888 and KIC~10273246) and two G-type stars (KIC~11395018 and KIC~10920273). We refer to \citet{2011ApJ...733...95M} and \citet{2011arXiv1108.3807C} for a detailed analysis of these stars. We present here the results of the analysis of two of these solar-type stars, KIC~11395018 and KIC~11234888, that have been observed for more than eight months.

\section{Analysis of the {\it Kepler} targets}
%%-------------------------

In each observation quarter (of $\sim$~3 months), a few hundreds of stars are observed by {\it Kepler}  in  short cadence (58.85~s) \citep{2010ApJ...713L.160G}. 
The {\it Kepler} Science Office provides light curves that are corrected for instrumental effects and that are optimized for exoplanet transits search \citep{2010ApJ...713L..87J}. However, it can happen that the low-frequency signals of some stars are filtered out. Thus, within the Kepler Asteroseismic Scientific Consortium \cite[KASC,][]{2010AN....331..966K}, the Working Group~\#~1 (which focuses on oscillations in main-sequence solar-type stars) developed their own method for the corrections. The raw light curves are corrected for three types of instrumental perturbations: outliers, jumps, and drifts, following the methods described in \citet{2011MNRAS.414L...6G}. The data of each {\it Kepler} quarter are concatenated after equalizing their mean values by fitting a 6th order polynomial to each segment. To remove the low-frequency instrumental trends, a high-pass filter is applied. 

Several teams \citep{2009CoAst.160...74H, 2009A&A...508..877M, 2009A&A...506..435R,2010MNRAS.408..542C,2010MNRAS.402.2049H,2010A&A...511A..46M} analyzed the light curves to retrieve the global parameters of the p modes. We were able to measure the mean large frequency separation $\langle \Delta \nu \rangle$, the frequency of maximum oscillation power $\nu_{\rm max}$, the mean small separation $\langle \delta_{02} \rangle$, and the mean linewidth $\langle \Gamma \rangle$.  We obtained $\langle \Delta \nu \rangle$ = 47.76~$\pm$~0.99~$\mu$Hz and $\langle \Delta \nu \rangle$ = 41.74~$\pm$~0.94~$\mu$Hz for KIC~11395018 and KIC~11234888 respectively. We determined the mean values of the small separation, $\langle \delta_{02} \rangle$ = 4.12~$\pm$~0.035~$\mu$Hz and 2.38~$\pm$~0.19~$\mu$Hz respectively for KIC~11395018 and KIC~11234888, as well as the mean linewidth of the modes, $\langle \Gamma \rangle$ = 0.84~$\pm$~0.02~$\mu$Hz and 0.86~$\pm$~0.06~$\mu$Hz respectively.

We also fitted the granulation signal with a Harvey-like function \citep{1985ESASP.235..199H,2011arXiv1109.1194M}, yielding the granulation time scale, $\tau_{\rm gran}$ of $698\,\pm\,33$~s and $869\,\pm\,89$~s for KIC~11395018 and KIC~11234888 respectively.

To study the low-frequency region of the power spectrum, we used different high-pass filters so that the signal below 1~$\mu$Hz was conserved. We estimated the surface rotation period to 36 days for KIC~11395918. The data of KIC~11234888 were noisier than the rest of the data. So we only considered  the last 7 months of data. Several peaks were present at low frequency suggesting that the surface of that star has a differential rotation (between 19 and 27 days).  

In addition, several teams fitted the individual modes using different methods. The most common approach adopted was the maximum likelihood estimator method  \citep[e.g.][]{AppGiz1998}. Another group used Bayesian Markov Chain Monte Carlo algorithms \citep[MCMC; e.g.][]{2011A&A...527A..56H}. Finally a smaller group did not fit the individual modes but looked for the highest peaks in the power spectrum \citep[e.g.][]{2008ApJ...676.1248B}. Applying the method described in \citet{2011ApJ...733...95M}, we selected the minimal and maximal lists of frequencies that provide very tight constraints to model the stars.

%%
%% Example of two figures side by side
%%
\begin{figure}[ht!]
 \centering
 \includegraphics[width=6cm,clip, angle=90]{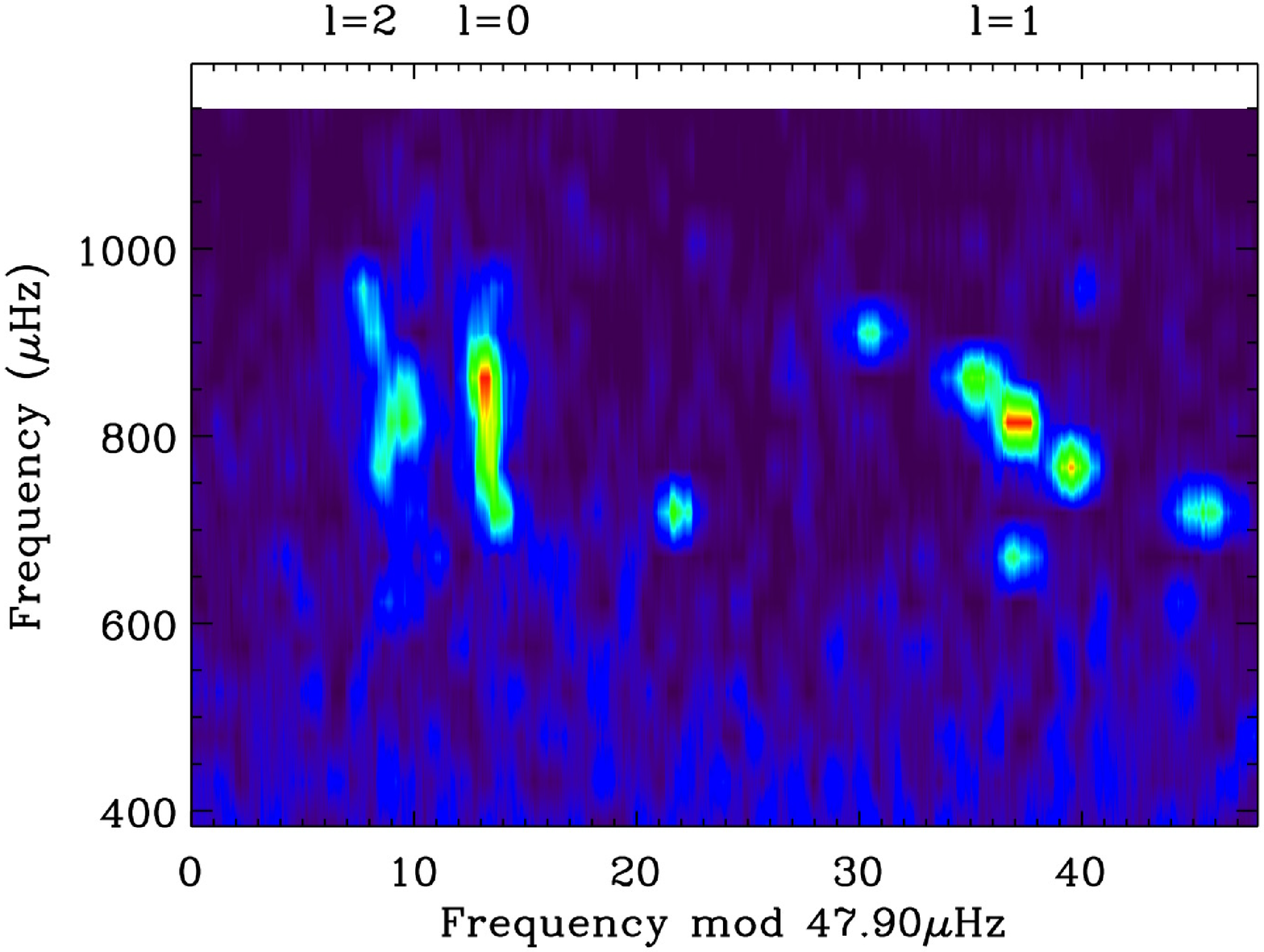}%      
 \includegraphics[width=6cm,clip, angle=90]{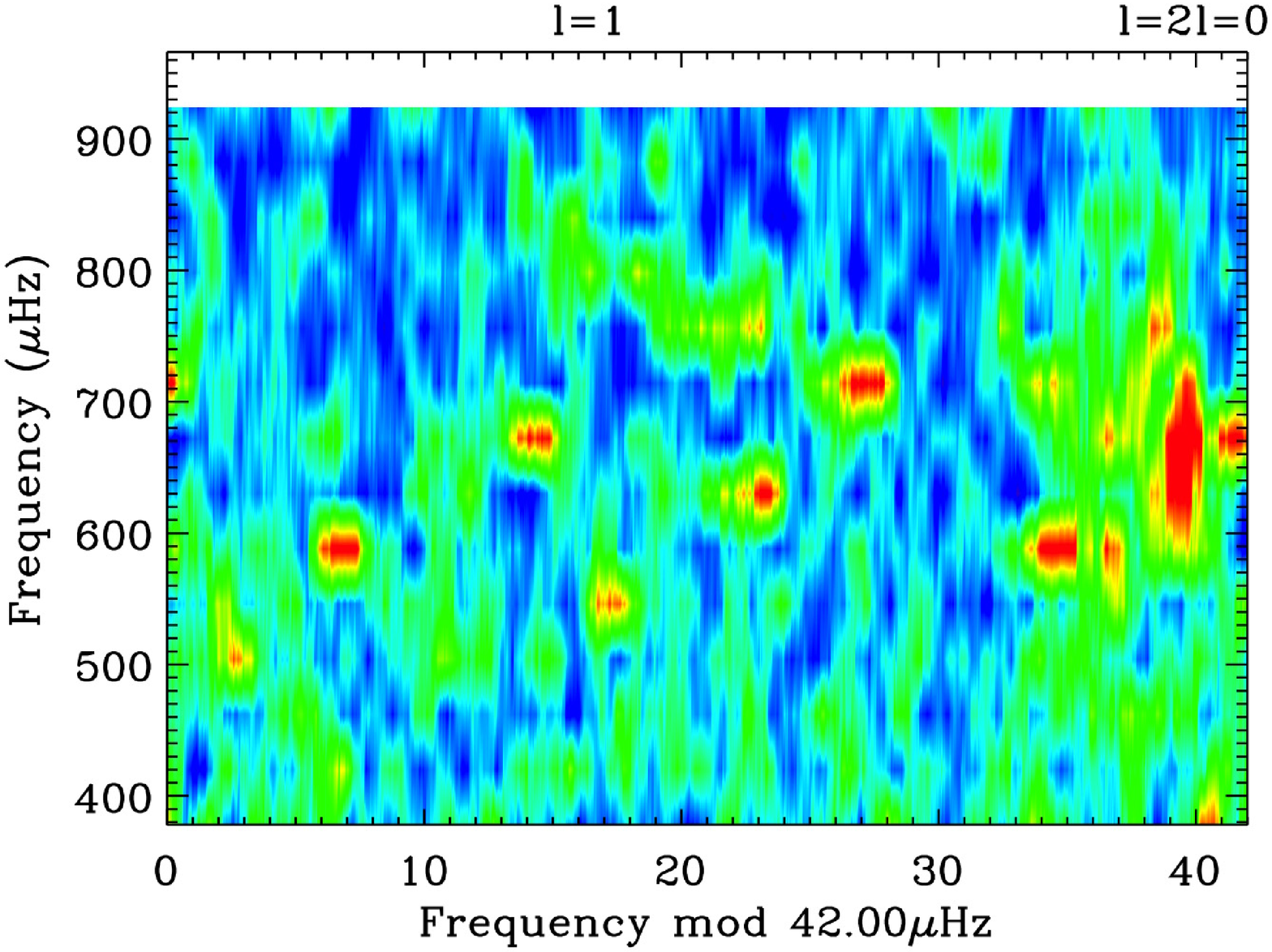}      
%% Note the ABSENCE of the extension .pdf , .eps or .ps  !
  \caption{{\bf Left:} Echelle diagram of 8 months of data for KIC~11395918. {\bf Right:} Echelle diagram of 8 months of data for the star KIC~11234888. }
  \label{mathur:fig1}
\end{figure}

Figure~\ref{mathur:fig1} shows the \'echelle diagrams of the two stars. This diagram is computed as follows: we take segments in the power spectrum of width approximately $\langle \Delta \nu \rangle$ and stack them one on top of each other. The left panel shows the ridges $\ell$\,=\,0, 1, and 2 for KIC~1139501. We can also note the presence of an avoided crossing, where an $\ell$~=~1 is ``bumped'' because of the coupling between acoustic and gravity modes. The right panel shows that KIC~11234888 has several avoided crossings. The presence of the mixed modes is very important to constrain the age of the star \citep[see for example][]{2010ApJ...723.1583M}.

Finally, we were able to have a first estimate for the splitting and inclination angle of the star KIC~11395918 using the MCMC approach. It gave an estimation of the projected rotational splitting $\nu_s^* = \nu_s sin(i)$ (where $\nu_s$ is the splitting) of 0.29\,$\pm$\,0.06\,~$\mu$Hz and an inclination angle $i > 20^\circ$, with a confidence level of 68\,\%.

\section{Conclusions}
%%--------------------

We analyzed the power spectra of two solar-type stars observed by {\it Kepler} during approximately eight months. We obtained not only the first estimates of the p-mode global parameters but also the individual frequencies of the p modes. We identified 22 p modes in the range 600 to 1000~$\mu$Hz for KIC~11395018 and 16 p modes in the range 500 to 900~$\mu$Hz for KIC~11234888.

Using scaling relations \citep{2011A&A...529L...8K,2010ApJ...723.1607H,2011MNRAS.415.3539V}, we can have a first estimate of the mass and radius of the stars. For KIC~11395018, we obtain: $M$~=~1.25~$\pm$~0.24~$M_{\odot}$ and $R$~=~2.15~$\pm$~0.21~$R_{\odot}$, while for KIC~11234888, we have $M$~=~1.33~$\pm$~0.26~$M_{\odot}$ and $R$~=~2.4$\pm$~0.24~$R_{\odot}$. 

These stars will be studied in more details using stellar modeling that combines the individual mode frequencies and the atmospheric parameters to retrieve precise estimates of the stellar parameters \citep[][Brand\~ao et al., in prep.; Do\u{g}an et al., in prep.]{creevey2011}.

This is the beginning of a new era as we are now able to measure the frequencies of p modes in a few hundreds of solar-type stars. This allow us to start testing and improving the stellar modeling codes with a very large sample of stars for the first time and to better understand stellar evolution. 

% Optional acknowledgements
% -------------------------
\begin{acknowledgements}
Funding for this Discovery mission is provided by NASAs Science Mission Directorate. The author wishes to thank the entire Kepler team, without whom these results would not be possible. The author also thanks the members of the KASC Working Group 1, all funding councils and agencies that have supported the activities of KASC Working Group 1, and the International Space Science Institute (ISSI). NCAR is supported by the National Science Foundation. 
\end{acknowledgements}

\bibliographystyle{aa}  % A&A bibliography style file (aa.bst)
%\bibliography{sf2a-template} % your references in file: Yourfile.bib
\bibliography{/Users/Savita/Documents/BIBLIO_sav.bib}

\begin{thebibliography}{37}
\expandafter\ifx\csname natexlab\endcsname\relax\def\natexlab#1{#1}\fi

\bibitem[{{Appourchaux} {et~al.}(2010){Appourchaux}, {Belkacem}, {Broomhall},
  {Chaplin}, {Gough}, {Houdek}, {Provost}, {Baudin}, {Boumier}, {Elsworth},
  {Garc{\'{\i}}a}, {Andersen}, {Finsterle}, {Fr{\"o}hlich}, {Gabriel}, {Grec},
  {Jim{\'e}nez}, {Kosovichev}, {Sekii}, {Toutain}, \&
  {Turck-Chi{\`e}ze}}]{2010A&ARv..18..197A}
{Appourchaux}, T., {Belkacem}, K., {Broomhall}, A.-M., {et~al.} 2010, \aapr,
  18, 197

\bibitem[{{Appourchaux} {et~al.}(1998){Appourchaux}, {Gizon}, \&
  {Rabello-Soares}}]{AppGiz1998}
{Appourchaux}, T., {Gizon}, L., \& {Rabello-Soares}, M.-C. 1998, \aaps, 132,
  107

\bibitem[{{Arentoft} {et~al.}(2008){Arentoft}, {Kjeldsen}, {Bedding}, {Bazot},
  {Christensen-Dalsgaard}, {Dall}, {Karoff}, {Carrier}, {Eggenberger},
  {Sosnowska}, {Wittenmyer}, {Endl}, {Metcalfe}, {Hekker}, {Reffert}, {Butler},
  {Bruntt}, {Kiss}, {O'Toole}, {Kambe}, {Ando}, {Izumiura}, {Sato}, {Hartmann},
  {Hatzes}, {Bouchy}, {Mosser}, {Appourchaux}, {Barban}, {Berthomieu},
  {Garcia}, {Michel}, {Provost}, {Turck-Chi{\`e}ze}, {Marti{\'c}}, {Lebrun},
  {Schmitt}, {Bertaux}, {Bonanno}, {Benatti}, {Claudi}, {Cosentino}, {Leccia},
  {Frandsen}, {Brogaard}, {Glowienka}, {Grundahl}, \&
  {Stempels}}]{2008ApJ...687.1180A}
{Arentoft}, T., {Kjeldsen}, H., {Bedding}, T.~R., {et~al.} 2008, \apj, 687,
  1180

\bibitem[{{Baglin} {et~al.}(2006){Baglin}, {Auvergne}, {Boisnard}, {Lam-Trong},
  {Barge}, {Catala}, {Deleuil}, {Michel}, \& {Weiss}}]{2006cosp...36.3749B}
{Baglin}, A., {Auvergne}, M., {Boisnard}, L., {et~al.} 2006, in COSPAR, Plenary
  Meeting, Vol.~36, 36th COSPAR Scientific Assembly, 3749

\bibitem[{{Ballot} {et~al.}(2011){Ballot}, {Gizon}, {Samadi}, {Vauclair},
  {Benomar}, {Bruntt}, {Mosser}, {Stahn}, {Verner}, {Campante},
  {Garc{\'{\i}}a}, {Mathur}, {Salabert}, {Gaulme}, {R{\'e}gulo}, {Roxburgh},
  {Appourchaux}, {Baudin}, {Catala}, {Chaplin}, {Deheuvels}, {Michel}, {Bazot},
  {Creevey}, {Dolez}, {Elsworth}, {Sato}, {Vauclair}, {Auvergne}, \&
  {Baglin}}]{2011A&A...530A..97B}
{Ballot}, J., {Gizon}, L., {Samadi}, R., {et~al.} 2011, \aap, 530, A97

\bibitem[{{Beck} {et~al.}(2011){Beck}, {Bedding}, {Mosser}, {Stello}, {Garcia},
  {Kallinger}, {Hekker}, {Elsworth}, {Frandsen}, {Carrier}, {De Ridder},
  {Aerts}, {White}, {Huber}, {Dupret}, {Montalb{\'a}n}, {Miglio}, {Noels},
  {Chaplin}, {Kjeldsen}, {Christensen-Dalsgaard}, {Gilliland}, {Brown},
  {Kawaler}, {Mathur}, \& {Jenkins}}]{2011Sci...332..205B}
{Beck}, P.~G., {Bedding}, T.~R., {Mosser}, B., {et~al.} 2011, Science, 332, 205

\bibitem[{{Bedding} {et~al.}(2011){Bedding}, {Mosser}, {Huber},
  {Montalb{\'a}n}, {Beck}, {Christensen-Dalsgaard}, {Elsworth},
  {Garc{\'{\i}}a}, {Miglio}, {Stello}, {White}, {De Ridder}, {Hekker}, {Aerts},
  {Barban}, {Belkacem}, {Broomhall}, {Brown}, {Buzasi}, {Carrier}, {Chaplin},
  {di Mauro}, {Dupret}, {Frandsen}, {Gilliland}, {Goupil}, {Jenkins},
  {Kallinger}, {Kawaler}, {Kjeldsen}, {Mathur}, {Noels}, {Aguirre}, \&
  {Ventura}}]{2011Natur.471..608B}
{Bedding}, T.~R., {Mosser}, B., {Huber}, D., {et~al.} 2011, \nat, 471, 608

\bibitem[{{Bonanno} {et~al.}(2008){Bonanno}, {Benatti}, {Claudi}, {Desidera},
  {Gratton}, {Leccia}, \& {Patern{\`o}}}]{2008ApJ...676.1248B}
{Bonanno}, A., {Benatti}, S., {Claudi}, R., {et~al.} 2008, \apj, 676, 1248

\bibitem[{{Borucki} {et~al.}(2010){Borucki}, {Koch}, {Basri}, {Batalha},
  {Brown}, {Caldwell}, {Caldwell}, {Christensen-Dalsgaard}, {Cochran},
  {DeVore}, {Dunham}, {Dupree}, {Gautier}, {Geary}, {Gilliland}, {Gould},
  {Howell}, {Jenkins}, {Kondo}, {Latham}, {Marcy}, {Meibom}, {Kjeldsen},
  {Lissauer}, {Monet}, {Morrison}, {Sasselov}, {Tarter}, {Boss}, {Brownlee},
  {Owen}, {Buzasi}, {Charbonneau}, {Doyle}, {Fortney}, {Ford}, {Holman},
  {Seager}, {Steffen}, {Welsh}, {Rowe}, {Anderson}, {Buchhave}, {Ciardi},
  {Walkowicz}, {Sherry}, {Horch}, {Isaacson}, {Everett}, {Fischer}, {Torres},
  {Johnson}, {Endl}, {MacQueen}, {Bryson}, {Dotson}, {Haas}, {Kolodziejczak},
  {Van Cleve}, {Chandrasekaran}, {Twicken}, {Quintana}, {Clarke}, {Allen},
  {Li}, {Wu}, {Tenenbaum}, {Verner}, {Bruhweiler}, {Barnes}, \&
  {Prsa}}]{2010Sci...327..977B}
{Borucki}, W.~J., {Koch}, D., {Basri}, G., {et~al.} 2010, Science, 327, 977

\bibitem[{{Bruntt} {et~al.}(2007){Bruntt}, {Su{\'a}rez}, {Bedding}, {Buzasi},
  {Moya}, {Amado}, {Mart{\'{\i}}n-Ruiz}, {Garrido}, {L{\'o}pez de Coca},
  {Rolland}, {Costa}, {Olivares}, \&
  {Garc{\'{\i}}a-Pelayo}}]{2007A&A...461..619B}
{Bruntt}, H., {Su{\'a}rez}, J.~C., {Bedding}, T.~R., {et~al.} 2007, \aap, 461,
  619

\bibitem[{{Campante} {et~al.}(2011){Campante}, {Handberg}, {Mathur},
  {Appourchaux}, {Bedding}, {Chaplin}, {Garc{\'{\i}}a}, {Mosser}, {Benomar},
  {Bonanno}, {Corsaro}, {Fletcher}, {Gaulme}, {Hekker}, {Karoff}, {R{\'e}gulo},
  {Salabert}, {Verner}, {White}, {Houdek}, {Brand{\~a}o}, {Creevey}, {Do{\u
  g}an}, {Bazot}, {Christensen-Dalsgaard}, {Cunha}, {Elsworth}, {Huber},
  {Kjeldsen}, {Lundkvist}, {Molenda-{\.Z}akowicz}, {Monteiro}, {Stello},
  {Clarke}, {Girouard}, \& {Hall}}]{2011arXiv1108.3807C}
{Campante}, T.~L., {Handberg}, R., {Mathur}, S., {et~al.} 2011, ArXiv e-prints
  1108.3807

\bibitem[{{Campante} {et~al.}(2010){Campante}, {Karoff}, {Chaplin}, {Elsworth},
  {Handberg}, \& {Hekker}}]{2010MNRAS.408..542C}
{Campante}, T.~L., {Karoff}, C., {Chaplin}, W.~J., {et~al.} 2010, \mnras, 408,
  542

\bibitem[{{Chaplin} {et~al.}(2011){Chaplin}, {Kjeldsen},
  {Christensen-Dalsgaard}, {Basu}, {Miglio}, {Appourchaux}, {Bedding},
  {Elsworth}, {Garc{\'{\i}}a}, {Gilliland}, {Girardi}, {Houdek}, {Karoff},
  {Kawaler}, {Metcalfe}, {Molenda-{\.Z}akowicz}, {Monteiro}, {Thompson},
  {Verner}, {Ballot}, {Bonanno}, {Brand{\~a}o}, {Broomhall}, {Bruntt},
  {Campante}, {Corsaro}, {Creevey}, {Do{\u g}an}, {Esch}, {Gai}, {Gaulme},
  {Hale}, {Handberg}, {Hekker}, {Huber}, {Jim{\'e}nez}, {Mathur}, {Mazumdar},
  {Mosser}, {New}, {Pinsonneault}, {Pricopi}, {Quirion}, {R{\'e}gulo},
  {Salabert}, {Serenelli}, {Aguirre}, {Sousa}, {Stello}, {Stevens}, {Suran},
  {Uytterhoeven}, {White}, {Borucki}, {Brown}, {Jenkins}, {Kinemuchi}, {Van
  Cleve}, \& {Klaus}}]{2011Sci...332..213C}
{Chaplin}, W.~J., {Kjeldsen}, H., {Christensen-Dalsgaard}, J., {et~al.} 2011,
  Science, 332, 213

\bibitem[{{Creevey} {et~al.}(2011){Creevey}, {Do{\u g}an}, {Frasca},
  {Thygesen}, {Basu}, {Bhattacharya}, {Biazzo}, {Brand{\~a}o}, {Bruntt},
  {Mazumdar}, \& et~al.}]{creevey2011}
{Creevey}, O.~L., {Do{\u g}an}, G., {Frasca}, A., {et~al.} 2011, \apj,
  submitted

\bibitem[{{Garc{\'{\i}}a} {et~al.}(2011){Garc{\'{\i}}a}, {Hekker}, {Stello},
  {Guti{\'e}rrez-Soto}, {Handberg}, {Huber}, {Karoff}, {Uytterhoeven},
  {Appourchaux}, {Chaplin}, {Elsworth}, {Mathur}, {Ballot},
  {Christensen-Dalsgaard}, {Gilliland}, {Houdek}, {Jenkins}, {Kjeldsen},
  {McCauliff}, {Metcalfe}, {Middour}, {Molenda-Zakowicz}, {Monteiro}, {Smith},
  \& {Thompson}}]{2011MNRAS.414L...6G}
{Garc{\'{\i}}a}, R.~A., {Hekker}, S., {Stello}, D., {et~al.} 2011, \mnras, 414,
  L6

\bibitem[{{Garc{\'{\i}}a} {et~al.}(2009){Garc{\'{\i}}a}, {R{\'e}gulo},
  {Samadi}, {Ballot}, {Barban}, {Benomar}, {Chaplin}, {Gaulme}, {Appourchaux},
  {Mathur}, {Mosser}, {Toutain}, {Verner}, {Auvergne}, {Baglin}, {Baudin},
  {Boumier}, {Bruntt}, {Catala}, {Deheuvels}, {Elsworth}, {Jim{\'e}nez-Reyes},
  {Michel}, {P{\'e}rez Hern{\'a}ndez}, {Roxburgh}, \&
  {Salabert}}]{2009A&A...506...41G}
{Garc{\'{\i}}a}, R.~A., {R{\'e}gulo}, C., {Samadi}, R., {et~al.} 2009, \aap,
  506, 41

\bibitem[{{Garc{\'{\i}}a} {et~al.}(2007){Garc{\'{\i}}a}, {Turck-Chi{\`e}ze},
  {Jim{\'e}nez-Reyes}, {Ballot}, {Pall{\'e}}, {Eff-Darwich}, {Mathur}, \&
  {Provost}}]{2007Sci...316.1591G}
{Garc{\'{\i}}a}, R.~A., {Turck-Chi{\`e}ze}, S., {Jim{\'e}nez-Reyes}, S.~J.,
  {et~al.} 2007, Science, 316, 1591

\bibitem[{{Gilliland} {et~al.}(2010){Gilliland}, {Jenkins}, {Borucki},
  {Bryson}, {Caldwell}, {Clarke}, {Dotson}, {Haas}, {Hall}, {Klaus}, {Koch},
  {McCauliff}, {Quintana}, {Twicken}, \& {van Cleve}}]{2010ApJ...713L.160G}
{Gilliland}, R.~L., {Jenkins}, J.~M., {Borucki}, W.~J., {et~al.} 2010, \apjl,
  713, L160

\bibitem[{{Handberg} \& {Campante}(2011)}]{2011A&A...527A..56H}
{Handberg}, R. \& {Campante}, T.~L. 2011, \aap, 527, A56

\bibitem[{{Harvey}(1985)}]{1985ESASP.235..199H}
{Harvey}, J. 1985, in ESA Special Publication, Vol. 235, Future Missions in
  Solar, Heliospheric \& Space Plasma Physics, ed. E.~{Rolfe} \& B.~{Battrick},
  199

\bibitem[{{Hekker} {et~al.}(2010){Hekker}, {Broomhall}, {Chaplin}, {Elsworth},
  {Fletcher}, {New}, {Arentoft}, {Quirion}, \&
  {Kjeldsen}}]{2010MNRAS.402.2049H}
{Hekker}, S., {Broomhall}, A., {Chaplin}, W.~J., {et~al.} 2010, \mnras, 402,
  2049

\bibitem[{{Huber} {et~al.}(2010){Huber}, {Bedding}, {Stello}, {Mosser},
  {Mathur}, {Kallinger}, {Hekker}, {Elsworth}, {Buzasi}, {De Ridder},
  {Gilliland}, {Kjeldsen}, {Chaplin}, {Garc{\'{\i}}a}, {Hale}, {Preston},
  {White}, {Borucki}, {Christensen-Dalsgaard}, {Clarke}, {Jenkins}, \&
  {Koch}}]{2010ApJ...723.1607H}
{Huber}, D., {Bedding}, T.~R., {Stello}, D., {et~al.} 2010, \apj, 723, 1607

\bibitem[{{Huber} {et~al.}(2009){Huber}, {Stello}, {Bedding}, {Chaplin},
  {Arentoft}, {Quirion}, \& {Kjeldsen}}]{2009CoAst.160...74H}
{Huber}, D., {Stello}, D., {Bedding}, T.~R., {et~al.} 2009, Communications in
  Asteroseismology, 160, 74

\bibitem[{{Jenkins} {et~al.}(2010){Jenkins}, {Caldwell}, {Chandrasekaran},
  {Twicken}, {Bryson}, {Quintana}, {Clarke}, {Li}, {Allen}, {Tenenbaum}, {Wu},
  {Klaus}, {Middour}, {Cote}, {McCauliff}, {Girouard}, {Gunter}, {Wohler},
  {Sommers}, {Hall}, {Uddin}, {Wu}, {Bhavsar}, {Van Cleve}, {Pletcher},
  {Dotson}, {Haas}, {Gilliland}, {Koch}, \& {Borucki}}]{2010ApJ...713L..87J}
{Jenkins}, J.~M., {Caldwell}, D.~A., {Chandrasekaran}, H., {et~al.} 2010,
  \apjl, 713, L87

\bibitem[{{Kjeldsen} \& {Bedding}(2011)}]{2011A&A...529L...8K}
{Kjeldsen}, H. \& {Bedding}, T.~R. 2011, \aap, 529, L8

\bibitem[{{Kjeldsen} {et~al.}(2010){Kjeldsen}, {Christensen-Dalsgaard},
  {Handberg}, {Brown}, {Gilliland}, {Borucki}, \& {Koch}}]{2010AN....331..966K}
{Kjeldsen}, H., {Christensen-Dalsgaard}, J., {Handberg}, R., {et~al.} 2010,
  Astronomische Nachrichten, 331, 966

\bibitem[{{Mathur} {et~al.}(2010{\natexlab{a}}){Mathur}, {Garc{\'{\i}}a},
  {Catala}, {Bruntt}, {Mosser}, {Appourchaux}, {Ballot}, {Creevey}, {Gaulme},
  {Hekker}, {Huber}, {Karoff}, {Piau}, {R{\'e}gulo}, {Roxburgh}, {Salabert},
  {Verner}, {Auvergne}, {Baglin}, {Chaplin}, {Elsworth}, {Michel}, {Samadi},
  {Sato}, \& {Stello}}]{2010A&A...518A..53M}
{Mathur}, S., {Garc{\'{\i}}a}, R.~A., {Catala}, C., {et~al.}
  2010{\natexlab{a}}, \aap, 518, A53

\bibitem[{{Mathur} {et~al.}(2010{\natexlab{b}}){Mathur}, {Garc{\'{\i}}a},
  {R{\'e}gulo}, {Creevey}, {Ballot}, {Salabert}, {Arentoft}, {Quirion},
  {Chaplin}, \& {Kjeldsen}}]{2010A&A...511A..46M}
{Mathur}, S., {Garc{\'{\i}}a}, R.~A., {R{\'e}gulo}, C., {et~al.}
  2010{\natexlab{b}}, \aap, 511, A46

\bibitem[{{Mathur} {et~al.}(2011{\natexlab{a}}){Mathur}, {Handberg},
  {Campante}, {Garc{\'{\i}}a}, {Appourchaux}, {Bedding}, {Mosser}, {Chaplin},
  {Ballot}, {Benomar}, {Bonanno}, {Corsaro}, {Gaulme}, {Hekker}, {R{\'e}gulo},
  {Salabert}, {Verner}, {White}, {Brand{\~a}o}, {Creevey}, {Do{\u g}an},
  {Elsworth}, {Huber}, {Hale}, {Houdek}, {Karoff}, {Metcalfe},
  {Molenda-{\.Z}akowicz}, {Monteiro}, {Thompson}, {Christensen-Dalsgaard},
  {Gilliland}, {Kawaler}, {Kjeldsen}, {Quintana}, {Sanderfer}, \&
  {Seader}}]{2011ApJ...733...95M}
{Mathur}, S., {Handberg}, R., {Campante}, T.~L., {et~al.} 2011{\natexlab{a}},
  \apj, 733, 95

\bibitem[{{Mathur} {et~al.}(2011{\natexlab{b}}){Mathur}, {Hekker},
  {Trampedach}, {Ballot}, {Kallinger}, {Buzasi}, {Garcia}, {Huber}, {Jimenez},
  {Mosser}, {Bedding}, {Elsworth}, {Regulo}, {Stello}, {Chaplin}, {De Ridder},
  {Hale}, {Kinemuchi}, {Kjeldsen}, {Mullally}, \&
  {Thompson}}]{2011arXiv1109.1194M}
{Mathur}, S., {Hekker}, S., {Trampedach}, R., {et~al.} 2011{\natexlab{b}},
  ArXiv e-prints 1109.1194

\bibitem[{{Metcalfe} {et~al.}(2010){Metcalfe}, {Monteiro}, {Thompson},
  {Molenda-{\.Z}akowicz}, {Appourchaux}, {Chaplin}, {Do{\u g}an},
  {Eggenberger}, {Bedding}, {Bruntt}, {Creevey}, {Quirion}, {Stello},
  {Bonanno}, {Silva Aguirre}, {Basu}, {Esch}, {Gai}, {Di Mauro}, {Kosovichev},
  {Kitiashvili}, {Su{\'a}rez}, {Moya}, {Piau}, {Garc{\'{\i}}a}, {Marques},
  {Frasca}, {Biazzo}, {Sousa}, {Dreizler}, {Bazot}, {Karoff}, {Frandsen},
  {Wilson}, {Brown}, {Christensen-Dalsgaard}, {Gilliland}, {Kjeldsen},
  {Campante}, {Fletcher}, {Handberg}, {R{\'e}gulo}, {Salabert}, {Schou},
  {Verner}, {Ballot}, {Broomhall}, {Elsworth}, {Hekker}, {Huber}, {Mathur},
  {New}, {Roxburgh}, {Sato}, {White}, {Borucki}, {Koch}, \&
  {Jenkins}}]{2010ApJ...723.1583M}
{Metcalfe}, T.~S., {Monteiro}, M.~J.~P.~F.~G., {Thompson}, M.~J., {et~al.}
  2010, \apj, 723, 1583

\bibitem[{{Mosser} \& {Appourchaux}(2009)}]{2009A&A...508..877M}
{Mosser}, B. \& {Appourchaux}, T. 2009, \aap, 508, 877

\bibitem[{{Mosser} {et~al.}(2011){Mosser}, {Barban}, {Montalb{\'a}n}, {Beck},
  {Miglio}, {Belkacem}, {Goupil}, {Hekker}, {De Ridder}, {Dupret}, {Elsworth},
  {Noels}, {Baudin}, {Michel}, {Samadi}, {Auvergne}, {Baglin}, \&
  {Catala}}]{2011A&A...532A..86M}
{Mosser}, B., {Barban}, C., {Montalb{\'a}n}, J., {et~al.} 2011, \aap, 532, A86

\bibitem[{{Mosser} {et~al.}(2009){Mosser}, {Michel}, {Appourchaux}, {Barban},
  {Baudin}, {Boumier}, {Bruntt}, {Catala}, {Deheuvels}, {Garc{\'{\i}}a},
  {Gaulme}, {Regulo}, {Roxburgh}, {Samadi}, {Verner}, {Auvergne}, {Baglin},
  {Ballot}, {Benomar}, \& {Mathur}}]{2009A&A...506...33M}
{Mosser}, B., {Michel}, E., {Appourchaux}, T., {et~al.} 2009, \aap, 506, 33

\bibitem[{{Roxburgh}(2009)}]{2009A&A...506..435R}
{Roxburgh}, I.~W. 2009, \aap, 506, 435

\bibitem[{{Verner} {et~al.}(2011){Verner}, {Elsworth}, {Chaplin}, {Campante},
  {Corsaro}, {Gaulme}, {Hekker}, {Huber}, {Karoff}, {Mathur}, {Mosser},
  {Appourchaux}, {Ballot}, {Bedding}, {Bonanno}, {Broomhall}, {Garc{\'{\i}}a},
  {Handberg}, {New}, {Stello}, {R{\'e}gulo}, {Roxburgh}, {Salabert}, {White},
  {Caldwell}, {Christiansen}, \& {Fanelli}}]{2011MNRAS.415.3539V}
{Verner}, G.~A., {Elsworth}, Y., {Chaplin}, W.~J., {et~al.} 2011, \mnras, 415,
  3539

\bibitem[{{Walker} {et~al.}(2003){Walker}, {Matthews}, {Kuschnig}, {Johnson},
  {Rucinski}, {Pazder}, {Burley}, {Walker}, {Skaret}, {Zee}, {Grocott},
  {Carroll}, {Sinclair}, {Sturgeon}, \& {Harron}}]{2003PASP..115.1023W}
{Walker}, G., {Matthews}, J., {Kuschnig}, R., {et~al.} 2003, \pasp, 115, 1023

\end{thebibliography}

\end{document}